\documentclass[12pt]{article}

\evensidemargin\oddsidemargin

\usepackage{latexsym}

\newcommand{\set}[1]{\{#1\}}

\usepackage{latexsym}

\newtheorem{theo}{Theorem}

\newcommand{\N}{{\bf N}}
\newcommand{\Q}{{\bf Q}}

\newcommand{\x}{{\bf x}}
\newcommand{\y}{{\bf y}}

\newenvironment{Instruction}[2]%
{
\noindent {\bf #1 \hfill #2} \\[-1ex]
\begin{flushright}\begin{minipage}{6.0in}
   }%
{\end{minipage}
\end{flushright}}

\begin{document}

\title{Computing  A Glimpse of Randomness 
}


\author{Cristian S. Calude, Michael J. Dinneen, Chi-Kou Shu \\
\normalsize Department of Computer Science,
University of Auckland,\\
\normalsize Private Bag 92019, Auckland, New Zealand \\
\small {E-mails:} {\tt \{cristian,mjd,cshu004\}@cs.auckland.ac.nz}
}
\date{}  
\maketitle

\thispagestyle{empty}
\begin{abstract}

A Chaitin Omega number is the halting probability of a  universal Chaitin
 (self-delimiting
Turing) machine. Every Omega number  is  both {\it computably
enumerable} (the limit of a computable, increasing,
converging sequence of rationals) and {\it random} (its binary expansion
 is an algorithmic random sequence). In particular, every Omega number is
strongly non-computable.
 The aim of this paper is to describe a procedure, which combines
Java 
programming  and
mathematical proofs, for computing  the {\it exact values of the first 64
bits of a  Chaitin
Omega:} 
\begin{center}
$0000001000000100000110001000011010001111110010111011101000010000$.
\end{center}
 Full description of programs
and proofs will be given elsewhere.
\end{abstract}

\section{Introduction}
\label{section:introduction}
Any attempt  to compute the uncomputable or to decide the undecidable is
without doubt challenging,
but hardly new (see, for example,  Marxen and  Buntrock \cite{ marxen},
Stewart \cite{stewart}, Casti \cite{casti}).
This paper describes  a hybrid  procedure
(which combines Java 
programming  and
mathematical proofs) for computing  the {\it exact values of the first 64
bits of a concrete Chaitin
Omega number}, $\Omega_U$, the halting probability of the universal Chaitin
(self-delimiting Turing)
machine  $U$, see \cite{ch8}. Note that any Omega
number is not only
uncomputable, but random, making the computing
task even more demanding.

Computing  lower bounds for $\Omega_U$ is not difficult: we  just  generate
more and more
halting programs. Are the bits produced by such a procedure exact?  {\em
Hardly}.
If the first bit of the approximation happens to be 1, then sure, it is
exact. However, if the
provisional bit given by an approximation is 0, then, due to possible
overflows,
 nothing prevents
the first bit of $\Omega_U$ to
 be either 0 or 1. This
situation extends to other bits as well. Only  an initial  run of  1's
may give  exact values
for some  bits of $\Omega_U$.

 The paper is structured as follows. Section 2 introduces the
basic notation. Computably
enumerable (c.e.)
reals, random reals and c.e.{}  random reals are presented in Section 3.
Various
theoretical difficulties preventing the exact computation of any  bits of an
Omega number are
discussed in Section 4. The register machine model of Chaitin
\cite{ch8} is discussed in Section 5.
In section 6 we summarize our computational results concerning the halting
programs of up to 84
bits long for $U$. They give a lower bound for $\Omega_U$ which is proved to
provide the exact values of the
first 64 digits of $\Omega_U$ in Section 7.

Chaitin \cite{gregemail00} has pointed out that the self-delimiting Turing machine constructed in the preliminary version of this paper \cite{cds} is universal in the sense of Turing (i.e., it is capable to simulate any 
self-delimiting Turing machine), but  it  {\it  is not} universal in the sense of algorithmic information
theory because the ``price" of simulation is not bounded by an additive constant; hence, the
halting probability is not an Omega number (but a c.e.{} real with some properties close  to randomness). The construction presented in this paper {\it is} a
self-delimiting Turing machine.  Full details will appear  in \cite{shu}.

\section{Notation}
\label{section:notation}

We will use  notation that is standard in  algorithmic
information theory; we will assume  familiarity with  Turing
machine computations, computable and computably enumerable (c.e.) sets (see,
for example,    Bridges
\cite{dbBook}, Odifreddi \cite{odi}, Soare
\cite{soare},  Weihrauch
\cite{weihrauch}) and elementary algorithmic information
theory (see,
for example,  Calude \cite{cris}).

 By
$\N, \Q$    we denote  the set of nonnegative integers (natural numbers)
and rationals,  respectively. If $S$ is a finite set, then $\#S$ denotes the number of elements of $S$.
Let $\Sigma=\{0,1\}$ denote the binary alphabet.
Let $\Sigma^*$ be the set of  (finite) binary strings,
and $\Sigma^\omega$ the set of infinite binary sequences.
The length of a string $x$  is denoted by $|x|$.
 A subset $A$ of $\Sigma^*$ is  {\em prefix-free } if whenever $s$
and $t$ are in $A$ and $s $ is a prefix of $ t$, then $s = t$.

For a sequence $\x = x_0x_1 \cdots x_n \cdots \in \Sigma^\omega$ and an
nonnegative integer  $n\ge 1$, $\x(n)$ denotes
the initial segment of length $n$ of $\x$  and
$x_i$ denotes the $i$th digit of $\x$, i.e.   $\x(n)=x_0 x_1 \cdots
x_{n-1}\in
\Sigma^*$. 
Due to Kraft's inequality, for every prefix-free set  $A \subset \Sigma^*,$
$ \Omega_A = \sum_{s \in A} 2^{- |s|}$
 lies in the interval $[0,1]$. In fact $ \Omega_A$ is a probability:
 Pick, at random using the Lebesgue measure on $[0,1]$, a real $\alpha$
in the unit interval and
note that the probability that some initial prefix of the binary
expansion of $\alpha$ lies in the prefix-free set $A$ is  exactly
$ \Omega_A.$

Following Solovay
\cite{solovaymanu,solovay2k} we say that  $C$ is a {\em (Chaitin)}
(self-delimiting Turing) {\em machine}, shortly,  a {\em machine}, if $C$
is a Turing machine processing binary strings such that its
program set (domain)
$PROG_C=\{x\in\Sigma^* \mid C(x) \mbox{  halts}\}$
is   a prefix-free set of strings.
Clearly,
$PROG_C$ is  c.e.;
conversely, every prefix-free  c.e.{}  set  of strings is the
domain of some  machine.
 The {\em program-size complexity} of the string
$x\in\Sigma^*$ (relatively to $C$)
is  $H_C(x)=\min \{|y| \mid  y \in \Sigma^*, \ C(y)=x\}$,
where $\min \emptyset = \infty$.
A major result of algorithmic information theory is the following
invariance relation: we can effectively construct  a  machine $U$ (called
{\em
universal}\/) such that  for every  machine
$C$, there is a constant $c>0$ (depending upon $U$ and $C$) such that for
every
$x, y \in \Sigma^*$ with $C(x)=y$, there exists a string $x'\in\Sigma^*$
with
$U(x')=y$ ($U$ simulates $C$) and $|x'| \leq |x| + c$ (the overhead for
simulation is no larger
than an
additive constant). In complexity-theoretic terms,
 $H_U (x) \leq H_C (x) + c$.
Note that $PROG_U$ is  c.e.{}  but not computable.

If $C$ is a machine, then $\Omega_C = \Omega_{PROG_{C}}$ represents its
halting probability. When
$C=U$ is a universal machine, then its halting probability $\Omega_U$ is
called
a {\em Chaitin $\Omega$ number}, shortly, {\em $\Omega$ number}.

\section{Computably Enumerable and Random Reals}
\label{section:randomreals}

Reals will be written in binary, so we start by looking at
random binary sequences.
Two complexity-theoretic definitions can be used to define
random sequences
(see Chaitin \cite{chaitin75,ch00}):
an infinite sequence $\mathbf{x}$ is {\em Chaitin random} if there
is a
constant $c$ such that $H(\mathbf{x}(n))>n-c$, for every integer $n>0$,
or, equivalently,  $\lim
_{n\rightarrow\infty}H(\mathbf{x}(n))-n=\infty$.
Other {\it equivalent} definitions include Martin-L\" of
\cite{martin1,martin} definition using statistical tests ({\em Martin-L\" of
random sequences}), Solovay \cite{solovaymanu}
 measure-theoretic definition ({\em Solovay random sequences}) and
 Hertling and Weihrauch \cite{HW98icalp} topological approach to define
randomness  ({\em Hertling--Weihrauch random sequences}).
Independent proofs
of the equivalence  between Martin-L\" of and Chaitin definitions have been obtained
by Schnorr and Solovay, cf. \cite{ch8,gregemail01}.
In what follows we will simply call ``random'' a sequence satisfying one of
the above equivalent conditions. Their equivalence motivates the
following ``randomness hypothesis''(Calude \cite{crisglimpse}):
\emph{A sequence is ``algorithmically random'' if it satisfies one of the
above
equivalent conditions.}
Of course, randomness implies strong non-computability (cf., for example,
Calude \cite{cris}),
but the converse is false.

 {\em A
real
$\alpha$ is random if its binary expansion
$\x$ (i.e.   $\alpha = 0.\x$) is random.}  The choice of the binary base
does not
play any role, cf. Calude and J\"{u}rgensen \cite{cj}, Hertling and
Weihrauch
\cite{HW98icalp}, Staiger \cite{staiger}: {\em randomness is a property of
reals not of names of reals.}

 Following  Soare \cite{soare1}, a real $\alpha$ is called {\em   c.e.}
if there is a computable, increasing sequence of rationals
which converges ({\em not necessarily computably}) to $\alpha$.
We will start with  several characterizations of  c.e.{}  reals (cf. Calude,
Hertling,
Khoussainov and
 Wang \cite{CHKW98stacs}).
If $0.\y$ is the binary expansion of a real $\alpha$ with
infinitely many ones, then $\alpha= \sum_{n\in X_\alpha } 2^{-n-1}$,
where
$X_\alpha =\{i \mid y_i=1\}$.

\begin{theo}
\label{aboutrereals}
Let $\alpha$ be a real in $(0,1]$. The following conditions are equivalent:
\begin{enumerate}
\item There is a computable, nondecreasing sequence
 of rationals  which converges to $\alpha$.
\item The set $\{p \in\Q \mid p < \alpha\}$ of rationals
less than $\alpha$ is  c.e.
\item There is an infinite prefix-free  c.e.\ set
$A \subseteq \Sigma^*$ with $\alpha = \Omega_A$.
\item There is an infinite prefix-free computable set
$A \subseteq \Sigma^*$ with $\alpha = \Omega_A$.
\item There is a total computable function $f:\N^2 \rightarrow \{0,1\}$
such that
\begin{enumerate}
\item If for some $k,n$ we have $f(k,n)=1 $ and $f(k,n+1)=0$ then there is
an $l<k$ with $f(l,n)=0$ and $f(l,n+1)=1$.
\item We have: $k \in X_\alpha \iff \lim_{n\rightarrow \infty} f(k,n) = 1$.
\end{enumerate}
\end{enumerate}
\end{theo}

We note that following Theorem~\ref{aboutrereals}, 5), given a computable
approximation
of a c.e.{}  real $\alpha$ via a total computable function $f$, $k \in
X_\alpha
\iff \lim_{n\rightarrow \infty} f(k,n) = 1$;
the values of $f(k,n)$ may oscillate from 0 to 1 and back; we will not be
sure that they stabilized until
$2^{k}$ changes have occurred (of course, there need not be so many changes,
but in this case there is no guarantee
of the exactness of the value of the $k$th bit).

 Chaitin \cite{chaitin75} proved the following important result:

\begin{theo}
\label{omegaisrandom}
If $U$ is a universal  machine, then $\Omega_U$ is c.e. and random.
\end{theo}

The converse of Theorem~\ref{omegaisrandom} is also true: it has been proved
by
Slaman \cite{slaman} based on work reported in Calude, Hertling,
Khoussainov and Wang \cite{CHKW98stacs} (see also Calude and Chaitin
\cite{cscgjc}, Calude \cite{cerandTCS}, Downey \cite{rod}):
\begin{theo}
\label{summaryomega}
Let $\alpha\in (0,1)$. The following
conditions are equivalent:
\begin{enumerate}
\item[{\rm 1.}] The real $\alpha$ is c.e.{}  and random.
\item[{\rm 2.}] For some universal  machine $U$,
$\alpha=\Omega_U$.
\end{enumerate}
\end{theo}

\section{The First Bits of An Omega Number}

We start by noting that

\begin{theo} 
\label{halting}
Given the first $n$ bits of $\Omega_U$ one can decide whether
$U(x)$ halts or not on an arbitrary string  $x$ of length at most $n$.
\end{theo}

 The first 10,000 bits of $\Omega_U$ include a tremendous amount of
mathematical knowledge.
In Bennett's words \cite{bg}:

\begin{quote}

\em 

 [$\Omega$] embodies an enormous amount of wisdom in a very small space
\ldots
inasmuch as its first few thousands digits, which could be written on a
small piece of paper, contain the answers to more mathematical questions
than
could be written down in the entire universe.

\medskip

Throughout history  mystics and philosophers have sought
a compact key to universal wisdom, a finite formula or text which, when
known
and understood, would
provide the answer to every question.
The use of the  Bible, the
Koran and
the I Ching for divination and the tradition of
 the  secret books of Hermes Trismegistus, and the medieval
Jewish Cabala exemplify this belief or hope. Such sources
of universal wisdom are traditionally protected from casual
use by being hard to find, hard to understand when found, and
dangerous to use, tending to answer more  questions and deeper ones  than
the searcher wishes to  ask.
 The esoteric book is, like
God, 
simple yet undescribable. It is  omniscient, and
transforms all who know it \ldots  Omega is in many senses a cabalistic
number. It can be known of, but not known, through human reason. To know
it in detail, one would have to accept its uncomputable digit sequence on
faith, like words of a sacred text.

\end{quote}

It is worth noting that even if we get, by some kind of miracle, the
first 10,000 digits of $\Omega_U$, the task of solving the problems
whose answers are embodied in these bits is computable but unrealistically
difficult:
the time it takes to find all halting programs of length less than $n$ from
$0.\Omega_0 \Omega_2 \ldots \Omega_{n-1}$ grows faster than any computable
function of
$n$.

Computing some initial bits of an Omega number is even more difficult.
According to Theorem~\ref{summaryomega},
c.e.{}   random reals can be coded by
  universal machines through their  halting probabilities. How ``good"
or ``bad" are these names?
In \cite{chaitin75} (see also \cite{ch97,ch98}), Chaitin
proved the following:

\begin{theo}\label{chaitin}
Assume that $ZFC$\footnote{Zermelo set theory with choice.} is
arithmetically
sound.\footnote{That is, any theorem of arithmetic proved by $ZFC$ is
\emph{true}.}
Then, for every  universal machine $U$, $ZFC$ can determine
the value of only finitely many bits of $\Omega_U$.
\end{theo}

In fact one can
give a bound on the number of bits of
$\Omega_U$ which
$ZFC$ can determine; this bound can be explicitly formulated, but it {\em
is  not computable}. For example,
 in \cite{ch97}  Chaitin described, in a dialect of Lisp, a  universal
machine $U$ and a  theory $T$, and proved that $U$ can determine the value
of
at most $H(T) + 15,328$ bits of $\Omega_U$;  $H(T)$ is the program-size
complexity
of the theory $T$, an {\em uncomputable} number.

Fix a universal  machine $U$ and consider all statements of the form
\begin{equation}
\label{*}
{\mbox {\rm ``The }} \,  n^{th}   \, {\mbox {\rm
 binary digit of the expansion of}}  \, \Omega_U \,\,  {\mbox  {\rm is }}
\,  k",
\end{equation}
 for all $n \ge 0, k = 0,1$. How many theorems of the form (\ref{*}) can
$ZFC$ prove?
More precisely, is there a bound on the set of non-negative integers $n$
such that
$ZFC$ proves a theorem of the form (\ref{*})? From Theorem~\ref{chaitin} we
deduce
that $ZFC$ can prove only finitely many (true) statements of the form
(\ref{*}). This is
Chaitin information-theoretic version of G\" odel's incompleteness (see
\cite{ch97,ch98}):

\begin{theo}
\label{chaitinincompleteness}
If $ZFC$ is arithmetically sound and $U$ is a  universal machine, then
almost all true statements of the form (\ref{*}) are unprovable in $ZFC$.
\end{theo}

Again, a bound can be explicitly found, but  not effectively computed.
Of course, for every c.e.{}  random real  $\alpha$ we can construct a
universal
machine
$U$  such  that  $\alpha = \Omega_U$ and $ZFC$ is able to determine finitely
(but as many as we
want) bits of $\Omega_U$.

A   machine $U$  for which Peano Arithmetic can
prove its universality and 
$ZFC$ cannot determine more than the initial
block of 1 bits of the binary expansion of its halting probability,
$\Omega_U$,
will be called  {\em Solovay machine}.\footnote{Clearly, $U$
depends on $ZFC$.} To make things worse   Calude \cite{crissol}  proved the
following
result:

\begin{theo}
\label{cristh}
Assume that  $ZFC$ is arithmetically sound. Then, every c.e.{}~random real
is the halting probability of a Solovay machine.
\end{theo}

For example, if $\alpha \in (3/4, 7/8)$ is c.e.{} and random, then in the
worst case
$ZFC$ can determine its first two bits (11), but no more. For $\alpha \in
(0, 1/2)$
we obtained Solovay's Theorem \cite{solovay2k}:

\begin{theo}
\label{cris}
Assume that  $ZFC$ is arithmetically sound. Then, every c.e.{}\ random real
 $\alpha \in (0, 1/2)$
is the halting probability of a Solovay machine which cannot determine
any single bit of $\alpha$. No c.e.{}  random real $\alpha \in (1/2, 1)$ has
the above
property.
\end{theo}
 
The conclusion is that the worst fears discussed in the first section proved
to 
materialize: In general only the initial run of 1's (if any) can be exactly
computed.

\section{Register Machine  Programs}

First we start with   the
register machine model  used by  Chaitin  \cite{ch8}.  Recall that
any register machine has a finite number of registers, each of which may
contain an arbitrarily large non-negative integer. The list
of instructions is given below in two forms: our compact form and
its corresponding Chaitin \cite{ch8} version. The main difference between
Chaitin's implementation and ours is in the encoding: we use 7 bit codes
instead of 
8 bit codes. 

\bigskip

\begin{Instruction}{L: ?  L1}{(L:  GOTO  L1)}
This is an unconditional branch to L1. L1 is a label of some
instruction in the program of the register machine.
\end{Instruction}

 \begin{Instruction}{L:  $\wedge$ \ R  L1}{ (L:  JUMP  R  L1)}
Set the register R to be the label of the next instruction and go to the
instruction with label L1.
\end{Instruction}

 \begin{Instruction}{L: @  R  }{                                    (L:
GOBACK  R)}
Go to the instruction with a label which is in R. This instruction will be
used in conjunction with
the jump instruction to return from a subroutine.  The instruction is illegal (i.e., run-time error
occurs) if R has not been explicitly set to a valid label of an instruction in the program.
\end{Instruction}

\begin{Instruction}{L:  =  R1  R2  L1  }{                           (L:  EQ
R1  R2  L1)}
This is a  conditional branch.  The last 7 bits of register R1 are
compared with the last 7 bits of register R2.
If they are equal, then the execution continues at
the instruction with label L1.  If they are not equal, then execution
continues with the next instruction in sequential order. R2 may be replaced
by a constant which can be represented by a 7-bit ASCII code, i.e. a 
constant from 0 to 127.
\end{Instruction}

\begin{Instruction}{L:  \#  R1  R2  L1 }{                            (L:
NEQ  R1  R2  L1)}
This is a  conditional branch.  The last 7 bits of register R1 are
compared with the last 7 bits of register R2.
If they are not equal, then the execution
continues at the instruction with label L1.  If they are equal, then
execution continues with the next instruction in sequential order. R2 may be
replaced by a constant which can be represented by a 7-bit ASCII code, i.e. 
a constant from 0 to 127.
\end{Instruction}

\begin{Instruction}{L:  )  R      }{                                (L:
RIGHT  R)}
Shift register R right 7 bits, i.e.,
the last character in R is deleted.
\end{Instruction}

\begin{Instruction}{L:  (  R1  R2   }{                               (L:
LEFT  R1  R2)}
Shift register R1 left 7 bits, add to it the rightmost  7 bits of register
R2, and then shift register R2 right 7 bits.
The register R2 may be replaced by a constant  from 0 to 127.
\end{Instruction}

\begin{Instruction}{L:  \&  R1  R2    }{                            (L:  SET
R1  R2)}
The contents of register R1 is replaced by the contents of
register R2.  
R2 may be replaced by a constant from 0 to 127.
\end{Instruction}

\begin{Instruction}{L:  !  R   }{                                   (L:  READ
R)}
One bit is read into the  register R, so the numerical value of R
becomes either 0 or 1.  Any attempt to read past the last data-bit
results in a run-time error.
\end{Instruction}

\begin{Instruction}{L:  / }{                                         (L:
DUMP)}
All register names and their contents, as bit strings, are written out.
This instruction is also used for debugging.
\end{Instruction}

\begin{Instruction}{L:  \%  }{                                      (L:
HALT)}
Halts the execution.  This is the last instruction for each register machine
program.
\end{Instruction}

\bigskip

\if01
The alphabet of the register machine programs consists of standard ASCII 7
bit codes for 

\begin{itemize}
\item the above 11 special specification symbols, ?,  $\wedge$ , @, =, \#,
), (, \&, !, /, \% for
instructions, 
\item all lower case letters, a, b, \ldots, z,
\item the digits 0, 1, \ldots, 9,

\item the upper case letters X and  C (for constants),
\item the space character,
\end{itemize}

\noindent which totals 50 out of 128 possible ASCII codes.

\bigskip
\fi

A {\em register machine program}
 consists of finite list of labeled instructions from
the above list, with the restriction that the \verb{HALT{  instruction appears only
once, 
as the last instruction of the list. The data (a binary string) follows immediately
the \verb{HALT{  instruction.
The use of undefined variables is a run-time error.
A program not reading the whole data or attempting 
 to read past the last data-bit results in
 a run-time error. 
Because of the
position of the
\verb{HALT{  instruction  and the specific way  data is read, register machine programs are Chaitin machines.

To be more precise, we present a context-free grammar $G$ = $(N, \Sigma, P, S)$ in
Backus-Naur form which generates the register machine programs.

\bigskip
\newcommand{\NONTERM}[1]{\mbox{$\langle #1 \rangle$}}
\newcommand{\NonTerm}[1]{\NONTERM{\mbox{\sc #1}}}
\newcommand{\Ins}{\mbox{Ins}}
\newcommand{\GOTOins}{\NONTERM{?_{\Ins}}}
\newcommand{\JUMPins}{\NONTERM{\wedge_{\Ins}}}
\newcommand{\GOBACKins}{\NONTERM{@_{\Ins}}}
\newcommand{\EQins}{\NONTERM{=_{\Ins}}} 
\newcommand{\NEQins}{\NONTERM{\#_{\Ins}}}
\newcommand{\RIGHTins}{\NONTERM{)_{\Ins}}}
\newcommand{\LEFTins}{\NONTERM{(_{\Ins}}}
\newcommand{\SETins}{\NONTERM{\mbox{\&}_{\Ins}}}
\newcommand{\READins}{\NONTERM{!_{\Ins}}}
\newcommand{\DUMPins}{\NONTERM{/_{\Ins}}}
\newcommand{\HALTins}{\NONTERM{\%_{\Ins}}}
\newcommand{\RMins}{\NONTERM{\mbox{RMS}_{\Ins}}}
\newcommand{\Data}{\NonTerm{Data}}

\newcommand{\Label}{\NonTerm{Label}}
\newcommand{\Register}{\NonTerm{Register}}
\newcommand{\Constant}{\NonTerm{Constant}}
\newcommand{\SpecialSymbol}{\NonTerm{Special}}
\newcommand{\WS}{\NonTerm{Space}}
\newcommand{\Alpha}{\NonTerm{Alpha}}
\newcommand{\LS}{\NonTerm{LS}}

\noindent
(1) $N$ is the finite set of nonterminal variables:

{\small
\[ \begin{array}{rcl}
N & = & \set{S} \cup INST \cup TOKEN \\[2ex]
INST & = & \set{ \RMins, \GOTOins, \JUMPins, \GOBACKins, \EQins, \NEQins, \\
   &  &       \RIGHTins, \LEFTins, \SETins, \READins, \DUMPins, \HALTins} \\[2ex]
TOKEN & = & \set{\Data, \Label, \Register, \Constant,  \\
   &  &  \SpecialSymbol, \WS, \Alpha, \LS}\\
\end{array} \]
}

\bigskip

\noindent  
(2) $\Sigma$, the alphabet of the register machine programs, is a finite set of terminals, disjoint from $N$:

{\small
\[ \begin{array}{rcl}
\Sigma & = & \Alpha \cup \SpecialSymbol \cup \WS \cup \Constant\\
\Alpha & = & \set{a, b, c, \ldots, z}\\
\SpecialSymbol & = & \set{:, /, ?, \wedge, @, =, \#, ), (, \&, !, ?, \%}\\
\WS & = & \{\mbox{`space',`tab'}\}\\
\Constant & = & \set{d \mid 0 \leq d \leq 127} \\
\end{array} \]
}
\bigskip

\noindent
(3) $P$ (a subset of 
       $N\times(N\cup\Sigma)^{*}$) is the finite set of rules  (productions):

 {\small
\[ \begin{array}{rcl}
S & \rightarrow & \RMins^{*}\HALTins \Data\\  
\\
\Data & \rightarrow &  ( 0 | 1 )^{*}  \\
\\
\Label & \rightarrow & 0 \mid  ( 1 | 2 | \ldots | 9 ) ( 0 | 1 | 2 | \ldots | 9 )^{*}  \\
\\

\LS & \rightarrow & :\WS^{*}\\
\\
\Register & \rightarrow & \Alpha (\Alpha \cup ( 0 | 1 | 2 | \ldots | 9 ))^{*}\\
\\
\RMins &  \rightarrow & 
\GOTOins \mid  \JUMPins \mid \GOBACKins \mid  \EQins \mid  \NEQins \mid \\
   &  & \RIGHTins \mid  \LEFTins \mid \SETins \mid \READins \mid \DUMPins \\
\\
& & \texttt{(L: HALT)}\\
\HALTins & \rightarrow & \Label \LS \%\\ 
\\
& &  \texttt{(L: GOTO  L1)}\\
\GOTOins & \rightarrow & \Label \LS ? \WS^{*} \Label\\
\\
& &  \texttt{(L: JUMP  R  L1)}\\
\JUMPins & \rightarrow & \Label \LS \wedge \WS^{*} \Register \WS^{+} \Label\\ 
\\

& &  \texttt{(L: GOBACK  R)}\\
\GOBACKins & \rightarrow & \Label \LS @ \WS^{*} \Register \\
\end{array}
\] 
\[\begin{array}{rcl}

\\
& & \texttt{(L: EQ  R  0/127 L1  or  L: EQ R R2 L1)}\\
\EQins & \rightarrow & \Label  \LS = \WS^{*} \Register \WS^{+} \Constant
                              \WS^{+} \\
& & \Label \mid 
 \Label \LS = \WS^{*} \Register \WS^{+} \Register\\
& &  \WS^{+} \Label\\
\\
& & \texttt{(L: NEQ  R  0/127 L1  or  L: NEQ R R2 L1)}\\
\NEQins & \rightarrow & \Label  \LS \# \WS^{*} \Register \WS^{+} \Constant
                              \WS^{+}\\
& &  \Label \mid  \Label \LS \# \WS^{*} \Register \WS^{+}\\
& &  \Register \WS^{+} \Label\\
\\
& &   \texttt{(L: RIGHT  R)}\\
\RIGHTins  & \rightarrow & \Label \LS ) \WS^{*} \Register\\ 
\\
& & \texttt{(L: LEFT  R  L1)}\\
\LEFTins & \rightarrow & \Label \LS ( \WS^{*} \Register \WS^{+}\Constant \mid\\
                       & &    \Label \LS ( \WS^{*} \Register \WS^{+}\Register\\
\\
& & \texttt{(L: SET  R  0/127 or L: SET  R  R2)}\\
\SETins & \rightarrow & \Label \LS \& \WS^{*} \Register \WS^{+} \Constant \mid\\
    & &              \Label \LS \& \WS^{*} \Register \WS^{+} \Register\\ 
\\  
& &  \texttt{(L: READ  R)}\\
\READins & \rightarrow & \Label \LS ! \WS^{*} \Register\\ 
\\

& & \texttt{(L: DUMP)}\\

\DUMPins & \rightarrow & \Label \LS /\\
\end{array} \]
}
\bigskip

\noindent (4) $S \in N$ is the start symbol for the set of register machine programs.  

\bigskip

It is important to observe that the above construction is  {\it universal} in the sense of algorithmic information theory (see the discussion of the end of Section~1).  Register machine programs
are self-delimiting because the \verb{HALT{  instruction is at the end of any valid program.
Note that  the data, which immediately follows the \verb{HALT{  instruction, is read
bit by bit with no endmarker. This type of 
construction has been first programmed
in Lisp  by Chaitin \cite{ch8,ch00}.

To minimize the number of programs of a given length that need to be
simulated, we have used
``canonical programs" instead of general  register machines programs.
A {\em canonical program} is a  register machine program in which (1) labels
appear in increasing numerical
order starting with 0, (2) new register names appear in increasing
lexicographical order starting
from `a', (3) there are no leading or trailing spaces, (4) operands
are separated by a single space, (5) there is no space after labels
or operators, (6)
instructions are separated by a single space. Note that {\it for every
register machine program
there is a unique canonical program which is equivalent to it}, that is, both
programs have the same domain and produce the same output on a given input. If $x$ is a program and $y$ is its 
canonical program, then $|y| \leq |x|$.\\

Here is an example of  a canonical program: 
\medskip

\begin{verbatim}
    0:!a 1:^b 4 2:!c 3:?11 4:=a 0 8 5:&c 110 6:(c 101 7:@b 8:&c 101 
    9:(c 113 10:@b 11:%10
\end{verbatim}

\medskip

To facilitate the understanding of the code we rewrite the instructions with
 additional comments and spaces:

\medskip

\begin{tabular}{l l}
\verb|0:! a|  &      // read the first data bit into register a\\
\verb|1:^ b 4|        &   // jump to a subroutine at line 4\\
\verb|2:! c|              &  // on return from the subroutine call  c is written out\\
\verb|3:? 11|          &  // go to the halting instruction\\
\verb|4:= a 0 8|   &// the right most 7 bits are compared with 127; if they\\
                    &  // are equal, then go to label 8\\
\verb|5:& c `n'|       &  // else, continue here and\\
\verb|6:( c `e'|         & // store the character string  \verb|`ne'|  in register c\\
\verb|7:@ b|         &  // go back to the instruction with label 2 stored in\\
                    & // register b \\
\verb|8:& c `e'|       & // store the character string \verb|`eq'|  in register c \\
\verb|9:( c `q' |  & \\
\verb|10:@ b|        & \\
\verb|11:%|     &    // the halting instruction \\
\verb|10|      &    // the input data \\
\end{tabular}

\medskip

For optimization reasons, our particular implementation designates the first maximal sequence of \verb{SET/LET{ instructions as (static) register pre-loading instructions. 
We ``compress" these canonical programs by (a) deleting all labels,
spaces and the colon symbol with the first non-static instruction 
having an implicit label 0, (b) 
separating multiple operands by a single comma symbol, (c) replacing constants 
with their ASCII numerical values. 
The compressed format of the above 
program is

\begin{verbatim}
       !a^b,4!c?11=a,0,8&c,110(,c,101@b&,c,101(,c,113@b%10
\end{verbatim}

Note that compressed programs  are canonical programs because during the process of ``compression" everything remains the same except for the elimination of space.
Compressed programs use an alphabet with
49 symbols (including the halting character). The length is calculated as the sum
of the program length  and the data
length (7 times the number of characters). For example, the length of the above program is $7 \times (49 + 2) = 357$.

{\em For the remainder of this paper we will
be  focusing on compressed programs}.

\section{Solving the Halting Problem for Programs Up to 84  Bits }
A Java version interpreter for  register machine compressed programs has been implemented;
it imitates
Chaitin's universal machine in \cite{ch8}. This interpreter has been used to
test the Halting Problem for all  register machine programs of at most 84
bits long.
The results have been obtained according to the following
procedure:
\begin{enumerate}
\item Start by generating
all programs of 7 bits and test which of them stops.  All strings of length
7 which can
be extended to programs are considered
prefixes for possible halting programs of length 14 or longer; they will simply be
called 
{\em prefixes}. In general, all strings of
length
$n$  which can be extended to programs
are {\em prefixes} for possible
halting programs of length $n+7$ or longer. {\em Compressed prefixes} are
prefixes of compressed (canonical)  programs.
\item Testing the Halting Problem for programs of length $n\in\{7,14,21,
\ldots, 84\}$ was done
by  running all candidates (that is, programs of length $n$ which are
extensions of prefixes  of
length
$n-7$) for up to 100 instructions, and
proving that any generated program  which does not halt after running
100 instructions
never halts. For example, (uncompressed) programs that match the  regular expression 
\verb{"0:\^ a 5.* 5:\? 0"{ never halt on any input.
\end{enumerate}

\medskip

For example, the programs
\verb{"!a!b!a!b/
data. The  program \verb{"!a?1!b

\medskip

One would naturally want to know the shortest program that
halts with more than 100 steps.  If this program is larger
than 84 bits, then all of our looping programs never halt.
The trivial program with a sequence of 100 dump instructions 
runs for 101 steps but can we do better?  The answer
is yes.  The following family of programs $\{P_1, P_2, \ldots\}$ 
recursively count to $2^i$ but have linear growth in size.
The programs $P_1$ through $P_4$ are given below:\footnote{In all cases
the data length is zero.}
\begin{verbatim}
  /&a,0=a,1,5&a,1?2%
  /&a,0&b,0=b,1,6&b,1?3=a,1,9&a,1?2%
  /&a,0&b,0&c,0=c,1,7&c,1?4=b,1,10&b,1?3=a,1,13&a,1?2%
  /&a,0&b,0&c,0&d,0=d,1,8&d,1?5=c,1,11&c,1?4=b,1,14&b,1?3=a,1,17&a,1?2%
\end{verbatim}

In order to create the program $P_{i+1}$ from $P_i$ only 4 instructions 
 are added, while updating `goto' labels.

The running time $t(i)$, excluding the halt instruction, of
program $P_i$ is found by the following recurrence:
$t(1)=6$, $t(i) = 2\cdot t(i-1) + 4$.
Thus, since $t(4)=86$ and $t(5)=156$, $P_5$ is the smallest program 
in this family to exceed 100 steps.  The size of $P_5$ is 86 bytes (602 bits),
which is smaller than the trivial dump program of 707 bits. 
It is an open question on what is the smallest program that halts after 100 
steps.  An hybrid program, given below, created by combining $P_2$ and 
the trivial dump programs is the smallest known.

\begin{verbatim}
   &a,0/&b,0/////////////////////=b,1,26&b,1?2=a,1,29&a,1?0%
\end{verbatim}

This program of 57 bytes (399 bits) runs for 102 steps.

Note that the problem of finding the smallest program
with the above property is undecidable (see \cite{ch98}).

The distribution  of halting compressed programs of up to
84 bits for  $U$, the universal machine processing compressed programs,
is presented in Table~1.  All binary strings
representing   programs have the
 length divisible by 7.

\bigskip
\if01
\begin{center}
\begin{tabular}{|r@{\hspace*{6ex}}|r@{\hspace*{6ex}}|r@{\hspace*{6ex}}|r@{\hspace*{6ex}}|}
\hline
\multicolumn{1}{|c|}{\bf  Program $+$  } &\multicolumn{1}{|c|}{ \bf Number of } & \multicolumn{1}{|c|}{\bf Number of  }
 &\multicolumn{1}{|c|}{ \bf Number of programs} \\
\multicolumn{1}{|c|}{\bf  data length }  &  \multicolumn{1}{|c|}{ \bf  halting programs}& 
\multicolumn{1}{|c|}{\bf looping programs}   &\multicolumn{1}{|c|}{ \bf with run-time errors}\\
\hline
7    &1 & 0   & 0   \\
14    & 1 & 0    & 0    \\
21    & 4 & 1    & 1    \\
28   & 8 & 3    & 2    \\
35   &57 & 19     & 17  \\
42    & 323 & 68    & 50    \\
 49     & 1187     &   425   &   407  \\   
 56  &  5452    &   2245   &   2176  \\
 63     &  23225    &  13119     & 12559    \\
 70     &   122331   &  72821    &   68450  \\
   77   & 624657     &   415109   & 389287    \\
84     &  3227131    & 2405533     &  2276776   \\
\hline
\end{tabular}
\fi 

\begin{center}
\begin{tabular}{|r@{\hspace*{6ex}}|r@{\hspace*{6ex}}||r@{\hspace*{6ex}}|r@{\hspace*{6ex}}|}
\hline
\multicolumn{1}{|c|}{\bf  Program plus  } &\multicolumn{1}{|c||}{ \bf Number of } & \multicolumn{1}{|c|}{\bf Program plus   }
 &\multicolumn{1}{|c|}{ \bf Number of } \\
\multicolumn{1}{|c|}{\bf  data length }  &  \multicolumn{1}{|c||}{ \bf  halting programs}& 
\multicolumn{1}{|c|}{\bf data length}   &\multicolumn{1}{|c|}{ \bf halting programs}\\
\hline
7    & 1 & 49   & 1012  \\
14    & 1 & 56    &  4382  \\
21    & 3 & 63   &  19164  \\
28   & 8 & 70    &   99785  \\
35   & 50 & 77     & 515279 \\
42    & 311  & 84    &  2559837  \\
\hline
\end{tabular}
\bigskip

Table 1.  Distribution of halting programs
\end{center}
\bigskip

\section{The First  64  Bits of $\Omega_U$}
 Computing all halting programs of up to 84
bits for $U$  seems to
give  the exact values of the first 84 bits of $\Omega_U$. False! To
understand the point
let's first ask ourselves whether
 the converse implication in Theorem~\ref{halting} is true? The answer is {\em
negative}.
Globally, if we can compute all bits of $\Omega_U$, then we can decide the
Halting
Problem for every program for $U$ and conversely. However, if we can solve
for $U$
the Halting
Problem for
all programs up to $N$ bits long we might not still get any exact value for
any bit
of $\Omega_U$ (less all values for the first $N$ bits). Reason: A large set of very long
halting programs can
contribute to the values of  more significant
 bits of the expansion of $\Omega_U$.

\medskip

So, to be able to compute the exact values of the first $N$ bits of
$\Omega_U$ we need to
be able to {\em prove} that longer programs do not affect  the
first $N$
bits of  $\Omega_U$.
And, fortunately, this is the case for our computation.
Due to our specific procedure for solving the Halting Problem discussed in Section 6,
any  compressed  halting   program of length $n$  has a  compressed  prefix  of length $n-7$.
This gives an upper bound for the number of possible compressed  halting programs of
length $n$.

\medskip

 Let
$\Omega_U^n$ be the approximation of $\Omega_U$ given by the summation of
all halting
programs of up
to  $n$ bits in length.

\medskip
Compressed prefixes are partitioned into 
two cases --- ones with an \verb{HALT{ (\%) instruction and ones without.
Hence, halting programs may have one of the following two forms: either ``$x \, y $ \verb{HALT{ $u$",
where $x$ is a prefix of length $k$ not containing \verb{HALT{, $y$ is a sequence of
instructions of length $n-k$ not containing \verb{HALT{  and $u$ is the  data of length $m \ge 0$, or ``$x\, u$", where  $x$ is a prefix of length $k$  containing
one occurrence of \verb{HALT{  followed by data (possibly empty) and $u$ is the  data of length $m \ge 1$. In both cases the prefix $x$  has been extended by at least one character. Accordingly, the ``tail" contribution to the value
of
$$\Omega_U =
\sum_{n=0}^{\infty}
\sum_{\{|w|=n, \, U(w)\,
\mbox{\footnotesize{halts}}\}} 2^{-|w|}$$ is bounded from above by the sum of the following two convergent series (which reduce to two independent sums of geometric progressions):

\[\sum_{m=0}^{\infty} \sum_{n=k}^{\infty} \underbrace{ \#\{x \mid  \mbox{prefix  }  \,  x \mbox{ not containing  \footnotesize{HALT}}, |x| = k\}}_{x} \cdot
\underbrace{ 47^{n-k}}_{y} \cdot \underbrace{ 1}_{\mbox{\footnotesize{HALT}}} \cdot \underbrace{ 2^m}_{u} \cdot  128^{-(n+m+1)},\]

\noindent and 
\[ \sum_{m=0}^{\infty} \underbrace{ \#\{x \mid  \mbox{prefix  }  \,  x \mbox{  containing \footnotesize{HALT}}, |x| = k\}}_{x} \cdot
\underbrace{ 2^m}_{u}  \cdot  128^{-(m+k)}.\]

 The number 47
comes from the fact that the alphabet has 48 characters and the last instruction before the data is \verb{HALT{  (\%).

\medskip

There are 402906842 prefixes not containing \verb{HALT{ and 1748380
prefixes  containing \verb{HALT{. Hence, the ``tail" contribution of all programs of length 91 or greater is
bounded by: 
\[\sum_{m=0}^{\infty} \sum_{n=13}^{\infty} 402906842 \cdot
47^{n-13} \cdot 2^m \cdot  128^{-(n+m+1)} +  
\sum_{m=0}^{\infty} 1748380 \cdot
2^m  \cdot  128^{-(m+13)}\] \\[-8ex]

\begin{eqnarray}
\label{xx}
& = & 402906842 \cdot  \frac{64}{128 \cdot 47^{13}} \cdot
\sum_{n=13}^{\infty} \left(\frac{47}{128}\right)^n + 1748380 \cdot \frac{1}{63 \cdot 128^{13}}  
< 2^{-68},
\end{eqnarray}

\noindent that is,  the first 68 bits of $\Omega_U^{84}$  ``may be" correct by
our
method. 
Actually we do not have
68 correct bits, but {\it only}  64 because adding a 1 to the 68th bit  may cause an
overflow up to the 65th bit. From (\ref{xx}) it follows that no other overflows may occur.
\bigskip

The following list presents the main results of the computation:

\medskip

\noindent $\Omega^{7}_U$  = 0.0000001\\
$\Omega^{14}_U$ = 0.00000010000001\\
$\Omega^{21}_U$ = 0.000000100000010000011\\
$\Omega^{28}_U$ = 0.0000001000000100000110001000\\
$\Omega^{35}_U$ = 0.00000010000001000001100010000110010\\
$\Omega^{42}_U$ = 0.000000100000010000011000100001101000110111\\
$\Omega^{49}_U$ = 0.0000001000000100000110001000011010001111101110100\\
$\Omega^{56}_U$ = 0.00000010000001000001100010000110100011111100101100011110\\
$\Omega^{63}_U$ = 
0.000000100000010000011000100001101000111111001011101100111011100\\
$\Omega^{70}_U$ = 
0.0000001000000100000110001000011010001111110010111011100111001111001001\\
$\Omega^{77}_U$ = 
0.0000001000000100000110001000011010001111110010111011101000001110000010\\
\phantom{$\Omega_{77}$ =}1001111\\
$\Omega^{84}_U$ = 
0.0000001000000100000110001000011010001111110010111011101000010000011110\\
\phantom{$\Omega_{84}$ =}11011011011101\\

\medskip
\noindent The exact bits are underlined in   the  84
approximation:

\medskip

\noindent $\Omega^{84}_U$ =
0.\underline{0000001000000100000110001000011010001111110010111011101000010000}011110
\phantom{$\Omega_{84}$ =}11011011011101\\

\noindent In summary, the first 64 exact bits of   $\Omega_U$ are:

\bigskip
\noindent 
0000001000000100000110001000011010001111110010111011101000010000

\section{Conclusions}
The computation described in this paper is the first attempt to compute some initial
exacts bits of a random real. The method, which combines programming with mathematical proofs, can be improved in many respects. However, 
due to the impossibility of testing that  long
looping programs never actually  halt (the undecidability of the Halting Problem),
the method is essentially  non-scalable.

As we have already mentioned, solving the Halting Problem for programs of up to $n$ bits might not be
enough to compute exactly the first $n$ bits of the halting probability. In our case, we have solved
the Halting Problem for programs of at most 84 bits, but  we have obtained only 64 exact initial
bits of the halting probability.

Finally, there is no contradiction between Theorem~\ref{cris} and the
main result of this paper. $\Omega$'s are halting probabilities of Chaitin universal machines,
and each $\Omega$ is the halting probability of an infinite number of such
machines. Among them, there are those (called Solovay machines in \cite{crissol})
which are in a sense  ``bad" as $ZFC$ cannot determine more than the initial
run of 1's of their halting probabilities.  But the same $\Omega$ can be defined as the halting
probability of a Chaitin universal machine which is not a Solovay machine,
so $ZFC$, if supplied with that different machine, may be able to compute more
(but always, as Chaitin proved,  only finitely many)  digits of the same $\Omega$. Such a machine has been used for the $\Omega$ discussed in this paper.

 The web site {\tt ftp://ftp.cs.auckland.ac.nz/pub/CDMTCS/Omega/} contains all
programs used for the computation as well as  all intermediate and final data files (3 giga-bytes in
gzip format).

\section*{Acknowledgement} We thank Greg Chaitin for pointing out an error in our previous
attempt to compute the first  bits of an Omega number \cite{cds}, for
continuous  advice and encouragement.

\end{document}